\documentclass[aps,pre,10pt,showpacs,floatfix,twocolumn,nofootinbib,a4paper]{revtex4-1}

\usepackage{amsmath}
\usepackage{amssymb}
\usepackage{amsfonts}
\usepackage{graphicx,color}
\usepackage{bm}
\usepackage{mathrsfs}
\usepackage{verbatim}
\usepackage{url}

\newcommand{\vctr}[1]{\bm{#1}}
\renewcommand{\tensor}[1]{\mathbf{#1}}
\newcommand{\diver}{\operatorname{div}}
\newcommand{\curl}{\operatorname{curl}}
\newcommand{\tr}{\operatorname{tr}}

\newcommand{\dd}{\operatorname{d}\!}

\newcommand{\sgrad}{\nabla_\mathrm{{\!s}}}
\newcommand{\sdiv}{\diver_\mathrm{{\!s}}\!}
\newcommand{\scurl}{\curl_\mathrm{{s}}\!}
\newcommand{\n}{\bm{n}}
\newcommand{\np}{\bm{n}_\perp}
\newcommand{\m}{\bm{m}}

\newcommand{\e}{\bm{e}}

\newcommand{\uv}{\bm{u}}

\newcommand{\trans}{^\mathsf{T}}

\newcommand{\A}{\mathbf{A}}

\newcommand{\id}{\mathbf{I}}
\newcommand{\Lift}{\mathbf{L}}
\newcommand{\Q}{\mathbf{Q}}
\newcommand{\M}{\mathbf{M}}
\newcommand{\Proj}{\mathbf{P}}
\newcommand{\sgradn}{\sgrad\n}

\newcommand{\surface}{\mathscr{S}}

\newcommand{\normal}{\bm{\nu}}
\newcommand{\ctens}{\sgrad\normal}

\newcommand{\rs}{\bm{r}\!_S}
\newcommand{\rdot}{\dot{\bm{r}}}
\newcommand{\zero}{\bm{0}}
\newcommand{\rsurf}{\bm{r}\!_\surface}
\newcommand{\gradh}{\nabla h}

\newlength{\irrl}
\newlength{\irrw}
\newcommand{\irr}[1]{
	\settowidth{\irrl}{\mbox{$\displaystyle #1$}}
	\setlength{\irrw}{0.12ex}
	\mbox{$\hspace{0.2em}
		\stackrel{
			\mbox{$\vphantom{\rule[-5\irrw]{\irrw}{6\irrw}}
				\rule[-4\irrw]{\irrw}{5\irrw}\hspace{-\irrw}
				\rule{\irrl}{\irrw}\hspace{-\irrw}
				\rule[-4\irrw]{\irrw}{5\irrw}$}}
		{\mbox{$\displaystyle #1$}}\hspace{0.2em}$}
}

\newcommand{\planenu}{\mathscr{T}_{\normal}}
\newcommand{\planez}{\mathscr{T}_z}
\newcommand{\ero}{\bm{e}_\rho}
\newcommand{\ep}{\bm{e}_\phi}
\newcommand{\et}{\bm{e}_\vartheta}
\newcommand{\hess}{\nabla^2h}
\newcommand{\vv}{\bm{v}}
\newcommand{\ctensp}{(\ctens)_\perp}
\newcommand{\patch}{\mathscr{A}}

\newcommand{\radial}{\bm{r}}
\newcommand{\mol}{\bm{\ell}}
\newcommand{\ex}{\bm{e}_x}
\newcommand{\ey}{\bm{e}_y}
\newcommand{\ez}{\bm{e}_z}
\newcommand{\er}{\bm{e}_\rho}
\newcommand{\vt}{\vartheta}
\newcommand{\vp}{\phi}

\begin{document}

\title{Lifting map for ordered surfaces}
\author{Leonid V. Mirantsev}
\affiliation{Institute of the Problems of Mechanical Engineering, Academy of Sciences of Russia, St. Petersburg  199178, Russia}
\author{Andr\'e M. Sonnet}
\affiliation{Department of Mathematics and Statistics, University of Strathclyde,
Livingstone Tower, 26 Richmond Street, Glasgow G1 1XH, Scotland}
\author{Epifanio G. Virga}
\affiliation{Dipartimento di Matematica, Universit\`a di Pavia, Via Ferrata 5, 27100 Pavia, Italy}

\date{\today}

\begin{abstract}
When a material surface is functionalized so as to acquire some type of order,
functionalization of which soft condensed matter systems have recently provided many interesting examples,
the modeller faces an alternative. Either the order is described on the curved, physical surface where it belongs,
or it is described on a flat surface that is unrolled as pre-image of the physical surface under
a suitable \emph{height} function. This paper proposes a general method that pursues the
latter avenue by \emph{lifting} whatever order tensor is deemed appropriate from a flat to a curved surface.
To produce a specific application, we specialize this method to nematic shells, for which it also provides a simple,
but convincing interpretation of the outcomes of some molecular-dynamics experiments on ellipsoidal shells.   
\end{abstract}

\maketitle

\section{Introduction}
Ordered material surfaces represent a new frontier of soft matter science. Be surface order induced by adding
a coating nematic film onto a colloidal particle, as is the case for nematic shells  \cite{nelson:toward},
or by subtracting material in almost a tailorly fashion, as is the case for graphenes \cite{yllanes:thermal},
it would be desirable to possess a general method that reduces the description of whatever order tensor is
involved on a curved surface to a parent order tensor defined on a flat surface.
This paper is designed to illustrate such a general method.

Our main mathematical tool to achieve this end, which is presented in Sec.~\ref{sec:lifting_tensor},
is the \emph{lifting} tensor, which acts on the unit vector fields entering the definition of a generic order tensor
in two space dimensions. As the name suggests, the lifting tensor maps a unit vector field defined on
a flat surface into a unit vector field everywhere tangent to a curved surface represented in terms
of the usual \emph{height} function. This tensor reveals itself as a viable tool to redo surface calculus
in an untraditional way, as shown in Sec.~\ref{sec:calculus}. 

To give a specific example of the potential applications of the general method proposed here,
we consider in Sec.~\ref{sec:shells} the case of nematic shells, for which the elastic energy functional is expressed,
albeit in a simplified instance, in terms of both a parent flat nematic director field and the height function
that represents the shell (or, more precisely, one of its halves). For ellipsoidal shells of revolution,
in Sec.~\ref{sec:ellipsoids}, we use our method to explain some molecular dynamics simulations that reach
equilibrium patterns with defect arrangements suggestive of an elastic competition between two antagonistic
director alignments. Although admittedly approximate, our account of such an antagonism is in a closed,
analytic form, and it is in a good quantitative agreement with the outcomes of the numerical experiments
performed with ellipsoids of revolution with different aspect ratios. 

Section~\ref{sec:conclusions} collects the conclusions of our study  and attempts to broaden our perspective
so as to encompass within the scope of our method the deformation of flexible surfaces with imprinted in-material order.
A technical appendix provides details on the sampling of axially symmetric surfaces
that was employed to interpret the molecular dynamics experiments in the language of order tensors
(and associated nematic directors).

\section{Lifting tensor}\label{sec:lifting_tensor}
An \emph{ordered} surface $\surface$ is a material surface embedded in three-dimensional space and endowed with an \emph{order tensor}.
The latter may be either a vector or a higher-rank tensor. For example, nematic shells, which  shall be considered in greater detail
in Secs.~\ref{sec:shells} and \ref{sec:ellipsoids} below, are characterized (in their \emph{director} description) by a unit vector field
$\n$ everywhere tangent to $\surface$. Alternatively, they can be described by a surface \emph{quadrupolar} tensor field $\Q$, that is,
a symmetric and traceless second-rank tensor field such that $\Q\normal=\zero$, where $\normal$ is the outer unit normal to $\surface$.
In this description, the nematic director $\n$ can be retraced as the eigenvector of $\Q$ with positive eigenvalue,
\begin{equation}\label{eq:Q_representation}
\Q=\lambda(\n\otimes\n-\np\otimes\np),\quad\lambda\geqq0,
\end{equation}
where $\np=\normal\times\n$ is the eigenvector of $\Q$ with negative eigenvalue.
Similarly, a more complicated structure is described by a surface \emph{octupolar} tensor field $\A$, that is,
a completely symmetric and traceless third-rank tensor field such that $\A\normal=\zero$, where  $\zero$ now denotes the null second-rank tensor.
As shown in \cite{virga:octupolar}, $\A$ can be represented as 
\begin{equation}\label{eq:A_representation}
\A=\lambda\irr{\n\otimes\n\otimes\n},
\end{equation}
where $\n$ is again a unit vector field everywhere tangent to $\surface$ and the superimposed bracket $\irr{\cdots}$ denotes the completely symmetric and traceless part of the tensor it surmounts.\footnote{As proven in \cite{gaeta:octupolar}, the simple representation of $\A$ in \eqref{eq:A_representation} is only valid in two space dimensions; already in three dimensions \eqref{eq:A_representation} is no longer valid.}  

The above examples illustrate how a generic order tensor on $\surface$ is intrinsically described by one unit tangent vector field
on $\surface$ (or possibly more) and one scalar field (or correspondingly more), which we conventionally denote by $\n$ and $\lambda$,
respectively.%
\footnote{Would a single unit vector and a single scalar fail to represent the surface order tensor under consideration,
one should resort to the generalized eigenvectors and eigenvalues, as discussed, for example, in \cite{chen:octupolar}.}
For surfaces $\surface$ that can be represented as graphs over a planar domain $S$, it would be interesting to represent
any unit vector field $\n$ as \emph{lifted} from a corresponding \emph{planar} unit vector field $\m$ defined on $S$.
In general, this would enable us to reduce any variational problem cast on $\surface$ for a surface order tensor to a corresponding
variational problem phrased on the domain $S$ for a planar order tensor. 
All geometric complications related to the non-planarity of $\surface$ will be explicitly absorbed into
the energy functional of the special problem under consideration. Once we learn how to replace an $\n$ with an $\m$,
we would have also learned how to construct the planar image  $(\lambda,\m)$ of any eigenpair of
a surface order tensor (of any prescribed rank) on $\surface$, as any \emph{eigenvalue} $\lambda$ is lifted
from $S$ onto  $\surface$ (as well as projected back) by simply preserving its value through composition with the function
representing $\surface$ over $S$. This is the strategy that we shall pursue to \emph{squeeze} onto a plane possibly elaborate
order textures on surfaces representable as graphs. In principle, it could also be extended to surfaces outside this restricted
class by use of an atlas of lifting maps. Here, for simplicity, we shall set aside this further complication. 

In the following, also in view of the application to nematic shells presented in Sec.~\ref{sec:shells}, we shall concentrate
on a single unit vector $\n$ everywhere tangent to $\surface$; we shall show how it is lifted from a planar, unit vector field $\m$
on a planar domain $S$.

Formally, we assume that $\surface$ can be represented as the graph of \emph{height function} $h$ on a domain $S$ in the plane
where $\m$ lies. For definiteness, we shall say that $S$ is a domain in the $x$-$y$-plane and that in the Cartesian coordinates
$(x,y,z)$ $\surface$ is described by $z=h(x,y)$.

Consider a curve $\rs$ in $S$ parametrized as
\begin{equation}\label{eq:curve_on_S}
\rs(s)=x(s)\e_x+y(s)\e_y,
\end{equation}
where $s$ is the arc-length and $\e_x$ and $\e_y$ are the coordinate unit vectors. Correspondingly, a curve $\rsurf$ is generated by lifting $\rs$ onto $\surface$,
\begin{equation}\label{eq:curve_on_surface}
\rsurf(s)=x(s)\e_x+y(s)\e_y+h(x(s),y(s))\e_z.
\end{equation}
By differentiating $\rsurf$ with respect to $s$ (and denoting this differentiation with a superimposed dot), we readily see from \eqref{eq:curve_on_surface} that
\begin{equation}\label{eq:curve_on_surface_dotted}
\rdot\!_\surface=\rdot\!_S+(\gradh\cdot\rdot\!_S)\e_z=(\id+\e_z\otimes\gradh)\rdot\!_S,
\end{equation}
where $\nabla$ is the gradient in two dimensions, so that 
\begin{equation}\label{eq:gradient_orthogonality_codition}
\gradh\cdot\e_z\equiv0.
\end{equation}

Letting the unit tangent $\rdot\!_S$ to $\rs$ coincide with the local value of a director field $\m$ on $S$ and setting
\begin{equation}\label{eq:lifting_tensor_definition}
\Lift:=\id+\e_z\otimes\gradh,
\end{equation} 
we obtain from \eqref{eq:gradient_orthogonality_codition} that the tangent to the lifted curve $\rsurf$ is oriented
along the vector $\m^\ast=\Lift\m$. Clearly, $\m^\ast$ need not be a unit vector, and so the \emph{lifted} director field $\n$
is defined by normalizing $\m^\ast$,
\begin{equation}\label{eq:n_lifted_definition}
\n:=\frac{\Lift\m}{\left|\Lift\m\right|}.
\end{equation}
We call $\Lift$ the \emph{lifting tensor} and we now explore some of its properties.

First, it follows from the general algebraic identity
\begin{equation}\label{eq:algebraic_identity}
\det(\id+\bm{a}\otimes\bm{b})=1+\bm{a}\cdot\bm{b}
\end{equation}
that, by \eqref{eq:gradient_orthogonality_codition},
\begin{equation}\label{eq:det_L}
\det\Lift=1,
\end{equation}
and so $\Lift$ is invertible and
\begin{equation}\label{eq:L_inverse}
\Lift^{-1}=\id-\e_z\otimes\gradh,
\end{equation}
which follows from the general property
\begin{equation}\label{eq:general_algebraic_property}
\left(\id+\bm{a}\otimes\bm{b}\right)^{-1}=\id-\frac{1}{1+\bm{a}\cdot\bm{b}}\bm{a}\otimes\bm{b}\quad\text{for}\quad\bm{a}\cdot\bm{b}\neq-1.
\end{equation}
Second, as a consequence of both \eqref{eq:det_L} and \eqref{eq:L_inverse}, the adjugate tensor $\Lift^\ast$ is given by 
\begin{equation}\label{eq:L_adjugate}
\Lift^\ast=\left(\Lift^{-1}\right)\trans=\id-\gradh\otimes\e_z,
\end{equation}
where $\trans$ denotes transposition.

Since $\m$ is a unit vector such that $\m\cdot\e_z\equiv0$, we can write
\begin{equation}\label{eq:Lm_norm}
|\Lift\m|^2=1+\mu^2,
\end{equation}
where we have set
\begin{equation}\label{eq:mu_definition}
\mu:=\gradh\cdot\m.
\end{equation}
By use of \eqref{eq:Lm_norm} and \eqref{eq:mu_definition}, we give $\n$ in \eqref{eq:n_lifted_definition} the following form
\begin{equation}\label{eq:n_formula}
\n=\frac{\m+\mu\e_z}{\sqrt{1+\mu^2}}.
\end{equation}
This relation can be easily inverted: we can obtain $\m$, if $\n$ is known, by projection on the $x$-$y$-plane,
\begin{equation}\label{eq:m_from_n_by_projection}
\m=\frac{\n-(\n\cdot\e_z)\e_z}{\sqrt{1-(\n\cdot\e_z)^2}}.
\end{equation}
This equation is valid under the assumption that $\n\cdot\e_z\neq\pm1$, an assumption which holds
for all $\n$
whenever the outward unit normal $\normal$ to $\surface$ satisfies the property
\begin{equation}\label{eq:geometric_property}
\normal\cdot\e_z\neq0.
\end{equation}

The lifting tensor $\Lift$ can also be used to express $\normal$ in terms of $\gradh$. If we orient $\surface$ so that $\normal\cdot\e_z\geqq0$, then $\normal$ can be obtained from the cross product of the lifted vectors $\Lift\e_x$ and $\Lift\e_y$:
\begin{equation}\label{eq:normal_representation}
\begin{split}
\normal=&\frac{\Lift\e_x\times\Lift\e_y}{|\Lift\e_x\times\Lift\e_y|}=\frac{\Lift^\ast(\e_x\times\e_y)}{|\Lift^\ast(\e_x\times\e_y)|}=\frac{\Lift^\ast\e_z}{|\Lift^\ast\e_z|}\\
=&\frac{\e_z-\gradh}{\sqrt{1+|\gradh|^2}},
\end{split}
\end{equation}
where use has also been made of \eqref{eq:gradient_orthogonality_codition} and \eqref{eq:L_adjugate}. It readily follows from \eqref{eq:normal_representation} that 
\begin{equation}\label{eq:geometric_property_satisfied}
\normal\cdot\e_z=\frac{1}{\sqrt{1+|\gradh|^2}},
\end{equation}
which makes \eqref{eq:geometric_property} automatically satisfied for any smooth $h$. 

Since $\m$ is essentially obtained from $\n$ through a projection onto the $x$-$y$-plane (followed by a normalization),
one could legitimately suspect that the lifting tensor $\Lift$ is a projection in disguise too (again, to within a normalization).
We shall see now that this is the case only in two special instances. Since $\n$ is tangent to $\surface$,
the only projection that could obtain it from $\m$ is $\Proj=\id-\normal\otimes\normal$. We thus seek the unit vectors $\uv$ on
the $x$-$y$-plane such that $\Lift$ and $\Proj$ agree on $\uv$ to within a normalization. This amounts to solving the equation
\begin{equation}\label{eq:parallelism_condition}
\Lift\uv\times\Proj\uv=\zero.
\end{equation}
Since $\e_z\cdot\uv=0$, it follows from \eqref{eq:normal_representation} that
\begin{equation}\label{eq:Pu}
\Proj\uv=\uv-\frac{1}{1+|\gradh|^2}(\gradh\cdot\uv)(\gradh-\e_z),
\end{equation}
whereas, by \eqref{eq:gradient_orthogonality_codition},
\begin{equation}\label{eq:Lu}
\Lift\uv=\uv+(\gradh\cdot\uv)\e_z.
\end{equation}
Making use of both \eqref{eq:Pu} and \eqref{eq:Lu} in \eqref{eq:parallelism_condition}, we readily arrive at
\begin{equation}\label{eq:resolving_formula}
\begin{split}
(\gradh\cdot\uv)\bigg\{\frac{1}{1+|\gradh|^2}\bigg[&\uv\times\e_z-\uv\times\gradh\\-(\gradh\cdot&\uv)\e_z\times\gradh\bigg]
+\e_z\times\uv\bigg\}=\zero.
\end{split}
\end{equation}
This equation is trivially satisfied for
\begin{equation}\label{eq:first_solution}
\gradh\cdot\uv=0.
\end{equation}
When $\gradh\cdot\uv\neq0$, since all vectors in the curly brackets of \eqref{eq:resolving_formula} lie on the $x$-$y$-plane, but $\uv\times\gradh$, which is parallel to $\e_z$, a necessary condition for \eqref{eq:resolving_formula} to hold is
\begin{equation}\label{eq:second_solution}
\uv\times\gradh=\zero.
\end{equation}
It is easily seen by direct inspection that \eqref{eq:second_solution} is also sufficient
to make \eqref{eq:resolving_formula} satisfied. We thus conclude that the lifting tensor
in \eqref{eq:lifting_tensor_definition} can be replaced by the projection $\Proj$
(appropriately rescaled) only when $\m$ is either parallel or perpendicular to the gradient
of the height function $h$. Although this may be the case in some special circumstances
(such as those considered in Sect.~\ref{sec:ellipsoids}), $\Lift$ and $\Proj$ cannot
in general be identified with one another (as they differ more than by a mere normalization).

\section{Surface calculus}\label{sec:calculus}
It is our aim in this section to review the fundamentals of calculus on a surface $\surface$ that can be expressed as the graph of a height function $h$ on a planar base set $S$. In particular, we shall show that the principal curvatures and the principal directions of curvature of $\surface$ can be easily obtained by solving an eigenvalue problem in the plane that contains $S$.

Our starting point will be the representation formula \eqref{eq:normal_representation}
for the outward normal $\normal$ to $\surface$. The first of its consequences
is that the area element $\dd a$ on $\surface$ is expressed by 
\begin{equation}\label{eq:area_element}
\dd a=|\Lift\e_x\times\Lift\e_y|\dd x\dd y=\sqrt{1+|\gradh|^2}\dd x\dd y.
\end{equation}
 
Since $h$ is a function defined on $S$, \eqref{eq:normal_representation} delivers $\normal$ in terms of $(x,y)$
at the point $(x,y,z)$ on $\surface$, where $z=h(x,y)$.
We now wish to compute in the same parametrization the curvature tensor $\sgrad\normal$ of $\surface$, where $\sgrad$ denotes the \emph{surface} gradient on $\surface$. The simplest way to do this is by differentiating $\normal$ along the curve $\rsurf$ parametrized in the arc-length of the base curve $\rs$. By the chain rule, \eqref{eq:normal_representation} gives
\begin{multline}\label{eq:normal_dot}
\dot\normal=-\Bigg(\frac{1}{1+|\gradh|^2}\normal\otimes\gradh\\+\frac{1}{\sqrt{1+|\gradh|^2}}\id\Bigg)(\nabla^2h)\Lift^{-1}\dot{\bm{r}}_\surface,
\end{multline}
where use has also been made of \eqref{eq:curve_on_surface_dotted}. Since, by definition, $\dot\normal=(\sgrad\normal)\dot{\bm{r}}_\surface$, for arbitrary curves $\rsurf$, it follows from \eqref{eq:normal_dot} that the curvature tensor $\sgrad\normal$ can be obtained from the restriction to the local tangent plane $\planenu$ to $\surface$ of a tensor expressed only in terms of the height function $h$, which we shall denote as  
\begin{equation}\label{eq:curvature_tensor_perp}
(\sgrad\normal)_\perp=-\left(\frac{1}{1+|\gradh|^2}\normal\otimes\gradh+\frac{1}{\sqrt{1+|\gradh|^2}}\id\right)(\nabla^2h),
\end{equation}
for convenience, 
implying that it acts on $\planenu$. To obtain \eqref{eq:curvature_tensor_perp},
\eqref{eq:L_inverse} has also been employed together with the identity $(\nabla^2h)\e_z\equiv\zero$.
The tensors $\ctens$ and $\ctensp$ would only differ on vectors along $\normal$,
so that we could also write
\begin{equation}
\sgrad\normal=(\sgrad\normal)_\perp(\id-\normal\otimes\normal).
\end{equation}

It is easily seen that $\ctensp$ duly maps $\planenu$ into itself. Indeed a generic vector of $\planenu$ is obtained by lifting a generic vector $\vv$ of the $(x,y)$ plane, which we shall denote in brief as $\planez$. Using \eqref{eq:curvature_tensor_perp}, \eqref{eq:normal_representation}, and \eqref{eq:lifting_tensor_definition}, and recalling that $\hess$ maps $\planez$ into itself, we arrive at the identity
\begin{equation}\label{eq:curvature_tensor_identity}
\normal\cdot\ctensp\Lift\vv=0.
\end{equation}
Similarly, since $\normal\cdot\Lift\uv=0$, for all $\uv\in\planez$, and, by \eqref{eq:lifting_tensor_definition} and \eqref{eq:normal_representation},
\begin{equation}
\Lift\trans\normal=\frac{1}{\sqrt{1+|\gradh|^2}}\e_z,\quad\Lift\trans\hess=\hess,
\end{equation}
we conclude that 
\begin{equation}\label{eq:curvature_bilinear_form}
\begin{split}
\Lift\uv\cdot\ctensp\Lift\vv&=-\frac{1}{\sqrt{1+|\gradh|^2}}\uv\cdot(\hess)\vv\\
&=\Lift\vv\cdot\ctensp\Lift\uv,
\end{split}
\end{equation} 
which shows that $\ctensp$ in \eqref{eq:curvature_tensor_perp} is a symmetric tensor of $\planenu$ into itself. Thus, there is an orthonormal basis $(\e_1,\e_2)$ in $\planenu$ such that 
\begin{equation}\label{eq:principal_basis_representation}
\ctens=\kappa_1\e_1\otimes\e_1+\kappa_2\e_2\otimes\e_2,
\end{equation}
where $\kappa_1$ and $\kappa_2$ are the \emph{principal curvatures} of $\surface$ and $(\e_1,\e_2)$ are the corresponding \emph{principal directions} of curvature.

This is a classical result, what is perhaps newer is our way of extracting from \eqref{eq:curvature_tensor_perp} simple,
compact formulas to express $\kappa_1$ and $\kappa_2$ in terms of the height function $h$ and
to lift  $(\e_1,\e_2)$ from a pair $(\uv_1,\uv_2)$ of (not necessarily orthonormal) vectors  of $\planez$.
Both these tasks are accomplished by seeking the critical points of the quadratic form $a(\uv)=\Lift\uv\cdot\ctensp\Lift\uv$
subject to the normalizing constraint $\Lift\uv\cdot\Lift\uv=1$. 
By \eqref{eq:curvature_bilinear_form}, this amounts to say that
\begin{equation}\label{eq:kappa_lambda}
\kappa_i=-\frac{1}{\sqrt{1+|\gradh|^2}}\lambda_i\quad i=1,2,
\end{equation}
where $\lambda_i$ are the critical values of the function $f$ defined on $\planez$ by
\begin{equation}\label{eq:f_definition}
f(\uv):=\frac{\uv\cdot(\hess)\uv}{\uv\cdot\M\uv}.
\end{equation} 
Here 
\begin{equation}\label{eq:A_definition}
\M:=\id+\gradh\otimes\gradh,
\end{equation}
as 
\begin{equation}
\Lift\uv\cdot\Lift\uv=\uv\cdot\Lift\trans\Lift\uv=\uv\cdot\left(\id+\gradh\otimes\gradh\right)\uv
\end{equation}
for all $\uv\in\planez$.

Since $\det\M=1+|\gradh|^2>0$, we can apply to $f$ the theory of simultaneous diagonalization of two quadratic forms (see, for example, p.~127 of \cite{biscari:mechanics}) and conclude that there are linearly independent vectors $(\uv_1,\uv_2)$ in $\planez$ such that 
\begin{equation}\label{eq:u_vectors}
\uv_i\cdot\M\uv_j=\delta_{ij}\quad\text{and}\quad(\hess-\lambda_i\M)\uv_i=\zero.
\end{equation}
Therefore, the $\lambda$'s that deliver the principal curvatures $\kappa$'s through \eqref{eq:kappa_lambda} are the roots of the secular equation
\begin{equation}\label{eq:secular_equation}
\det(\hess-\lambda\M)=0
\end{equation}
and the corresponding principal directions of curvature are
\begin{equation}\label{eq:lifted_e_i}
\e_i=\Lift\uv_i,
\end{equation}
which by \eqref{eq:u_vectors} duly satisfy the orthonormality condition $\e_i\cdot\e_j=\delta_{ij}$. 

To illustrate this method and its versatility, we apply it to the case where $\surface$ is a surface of revolution about the axis $\e_z$, which will be of further use in Sec.~\ref{sec:ellipsoids}. In this case, the height function $h$ depends only on the radial coordinate $\rho:=\sqrt{x^2+y^2}$, and $\gradh=h'\ero$, where $\ero$ is the radial unit vector and a prime denotes differentiation with respect to $\rho$. It is easily seen that
\begin{subequations}
\begin{eqnarray}
\Lift&=&\id+h'\e_z\otimes\ero,\label{eq:axis_symmetric_L}\\
\M&=&\id+h'^2\ero\otimes\ero,\label{eq:axis_symmetric_A}\\
\hess&=&h''\ero\otimes\ero+\frac{h'}{r}\ep\otimes\ep,\label{eq:axis_symmetric_Hess}
\end{eqnarray}
\end{subequations}
where $\ep=\e_z\times\ero$ is the tangential unit vector of polar coordinates. The eigenvalue problem \eqref{eq:secular_equation}
has then the solution 
\begin{equation}
\lambda_1=\frac{h''}{1+h'^2}\quad\text{and}\quad\lambda_2=\frac{h'}{\rho}
\end{equation}
with corresponding eigenvectors, normalized according to the first formula in \eqref{eq:u_vectors},
\begin{equation}
\uv_1=\frac{1}{\sqrt{1+h'^2}}\ero\quad\text{and}\quad\uv_2=\ep.
\end{equation}
Therefore the principal curvatures are
\begin{equation}\label{eq:axis_symmetric_curvatures}
\kappa_1=-\frac{h''}{(1+h'^2)^{3/2}}\quad\text{and}\quad\kappa_2=-\frac{h'}{\rho\sqrt{1+h'^2}},
\end{equation}
and the principal directions of curvature are designated by the unit vectors
\begin{equation}\label{eq:principalDirections}
\e_1=\frac{\ero+h'\e_z}{\sqrt{1+h'^2}},\quad\e_2=\ep.
\end{equation}
In particular, for a half-ellipsoid of revolution with semiaxes $a$ (along the symmetry axis) and $b$,
\begin{equation}\label{eq:ellipoid_h}
h(\rho)=a\sqrt{1-\frac{\rho^2}{b^2}},\quad0\leqq\rho\leqq b,
\end{equation}
and by \eqref{eq:axis_symmetric_curvatures}
\begin{equation}\label{eq:ellipsoid_curvatures}
\begin{split}
\kappa_1=\frac{\eta}{b}\left[1+(\eta^2-1)\frac{\rho^2}{b^2}\right]^{-3/2},\\
\kappa_2=\frac{\eta}{b}\left[1+(\eta^2-1)\frac{\rho^2}{b^2}\right]^{-1/2},
\end{split}
\end{equation}
where $\eta:=a/b$ is the ellipsoid's aspect ratio. These formulas agree with (44) and (45) of \cite{harris2006curvature}, which were obtained in the most traditional way. 

\section{Nematic shells}\label{sec:shells}
In this section we study the first, and perhaps most natural application of the lifting method presented in this paper.
This is the case of \emph{nematic shells}, rigid surfaces decorated with a nematic order induced by elongated molecules
gliding on a given surface under the constraint of remaining everywhere tangent to it, though in an arbitrary direction.
Such decorated surfaces with \emph{planar degenerate} anchoring may also be boundaries of colloidal particles, which,
at least for each of two fitting halves, can be described by our lifting method. This is a case  where a single director
$\m$ and its lifted correspondent $\n$ suffice to describe the ordered surface $\surface$ (or each half of the surface
bounding a colloidal particle).

Since the seminal paper of Nelson~\cite{nelson:toward}, much has been written about possible technological applications
of nematic shells, some perhaps more visionary than others. We refer the interested reader to a number of
reviews~\cite{lopez-leon:drops,lagerwall:new,mirantsev:defect,serra2016curvature,urbanski2017liquid} which also summarize
the most recent advances in this field, from both the theoretical and experimental approach. Here we shall be content
with showing how a mathematical theory for nematic shells based on a single director description can effectively be
phrased on a flat plane.

We shall take $\sgradn$ as the basic distortion measure, thus placing our model amid the \emph{extrinsic} elastic
theories of nematic shells, pioneered by \cite{helfrich:intrinsic} and further corroborated by \cite{selinger:monte},
which regard the intermolecular interactions, where the distortional energy is stored, as taking place in the
three-dimensional space surrounding the supporting surface. As shown in \cite{napoli2012extrinsic}, this view
leads one quite naturally to identify components of the elastic energy that couple orientation and curvature.
In \cite{sonnet:bistable}, we recently found in Levi-Civita's parallel transport a systematic way to separate
the purely distortional energy from the curvature counterpart imprinted in the surface, which was called the \emph{fossil energy}. 

Adopting the surface energy density $W(\n,\sgradn)$ arrived at from Frank's bulk
energy~\cite[Chap.~3]{virga:variational} through a standard dimension reduction \cite{napoli2012surface},
we write
\begin{equation}\label{eq:energy_density_W}
\begin{split}
W(\n,\sgradn)&=\frac12k_1(\sdiv\n)^2+\frac12k_2(\n\cdot\scurl\n)^2\\&
+\frac12k_3|\n\times\scurl\n|^2,
\end{split}
\end{equation}
where $k_i\geqq0$ are elastic constants with physical dimension of an energy, and $\sdiv\n$ and $\scurl\n$
denote the surface divergence and the surface curl of the nematic director subject to
\begin{equation}\label{eq:director_constraint}
\n\cdot\normal\equiv0\quad\text{on}\quad\surface.
\end{equation}
A noticeable case is obtained from \eqref{eq:energy_density_W} by setting $k_1=k_2=k_3=k>0$; this is known as the \emph{one-constant} approximation, which reduces $W$ to the form
\begin{equation}\label{eq:energy_density_one_constant}
W=\frac12k|\sgradn|^2,
\end{equation}
since  for a field $\n$ that obeys \eqref{eq:director_constraint}
\begin{equation}\label{eq:identity}
\tr(\sgradn)^2=(\tr\sgradn)^2.
\end{equation}

We proved in \cite{sonnet:bistable} that the fossil energy associated with \eqref{eq:energy_density_W} takes the form
\begin{equation}
W_0(\n,\ctens)=\frac12(k_2-k_3)|(\ctens)\n\times\n|^2+\frac12k_3|(\ctens)\n|^2.
\end{equation}
The distortional energy is then $W_\mathrm{d}:=W-W_0$, which can also be written explicitly as\footnote{With the aid of equations (22) and (28) of \cite{sonnet:bistable}.}
\begin{equation}
W_\mathrm{d}(\n,\sgradn)=\frac12k_1[\n_\perp\cdot(\sgradn)\n_\perp]^2+\frac12k_3[\n_\perp\cdot(\sgradn)\n]^2,
\end{equation}
where $\n_\perp:=\normal\times\n$.
While for $k_2\geqq k_3$, the fossil energy is minimized for $\n$ aligned with the principal direction of 
curvature having the smallest square curvature, for $k_2<k_3$ this is not necessarily the case.
As shown in \cite{sonnet:bistable}, in the latter case, the orientation preferred by $\n$ may also fail to be unique.

These conclusions are neatly arrived at when the principal curvatures and principal directions
of curvature of the surface $\surface$ are known explicitly. However, the situation is more intricate
when $\surface$ is represented by a generic height function $h$ and $\n$ is delivered by lifting $\m$
from $S$ unto $\surface$. Thus, here we first represent $W_0$ as a function of $h$ and $\m$.
To this end, we find it convenient to make use of the basis $(\uv_1,\uv_2)$ defined in the $x$-$y$-plane by \eqref{eq:u_vectors}, and to express $\m$ as\footnote{Note that $(\uv_1,\uv_2)$ is not necessarily an orthonormal basis.}
\begin{equation}\label{eq:m_director_components}
\m=m_1\uv_1+m_2\uv_2.
\end{equation}
By \eqref{eq:lifted_e_i} and \eqref{eq:u_vectors}, letting $\n=n_1\e_1+n_2\e_2$, we readily see that 
\begin{equation}\label{eq:n_components}
n_1=\frac{m_1}{\sqrt{m_1^2+m_2^2}},\quad n_2=\frac{m_2}{\sqrt{m_1^2+m_2^2}}.
\end{equation}
Combining \eqref{eq:principal_basis_representation} and \eqref{eq:kappa_lambda} with \eqref{eq:n_components}, we finally arrive at 
\begin{equation}\label{eq:W_0_m_parametrization}
\begin{split}
W_0&=\frac12\frac{1}{1+|\gradh|^2}\frac{1}{m_1^2+m_2^2}\\
&\times\left[(k_2-k_3)(\lambda_1-\lambda_2)^2\frac{m_1^2m_2^2}{m_1^2+m_2^2}+k_3(\lambda_1^2m_1^2+\lambda_2^2m_2^2) \right],
\end{split}
\end{equation}
where $\lambda_i$ are the roots of \eqref{eq:secular_equation}. Since, by \eqref{eq:n_components}, $\n$ is a unit vector whatever normalization is adopted for $\m$, $W_0$ can also be studied under the normalization $m_1^2+m_2^2=1$, which simplifies \eqref{eq:W_0_m_parametrization} considerably:
\begin{equation}\label{eq:W_0_m_parametrization_simplified}
\begin{split}
W_0&=\frac12\frac{1}{1+|\gradh|^2}\\
&\times\left[(k_2-k_3)(\lambda_1-\lambda_2)^2m_1^2m_2^2+k_3(\lambda_1^2m_1^2+\lambda_2^2m_2^2) \right]
\end{split}
\end{equation}
This equation formally parallels equation (30) of \cite{sonnet:bistable}, but it is explicitly written in the fixed $x$-$y$-plane,
instead of the variable  tangent plane $\planenu$. 
For a specific choice of $h$, the study of the minimizers of \eqref{eq:W_0_m_parametrization_simplified} would easily reveal the map of all  orientations preferred on $S$ by the fossil elastic energy.

Expressions for $W_\mathrm{d}$ similar to \eqref{eq:W_0_m_parametrization}, involving both the $m_i$ and their gradients,
could easily be given, but we found them far less concise and transparent than \eqref{eq:W_0_m_parametrization} and
omit them here.

It was remarked in \cite{sonnet:bistable} that the knowledge of the minimizers of $W_0$ does not in general suffice
to predict the state with minimum total elastic energy $W$, as the minimizers of $W_0$ can seldom be extended to
the whole surface $\surface$ without incurring distortional energy. So, as suggestive as the study of the minimizers of $W_0$ can be, it must be supplemented by the search for a global minimum. We shall perform such a search in the simple case of the one-constant approximation, also in view of the application of our method to the molecular dynamics simulations on ellipsoidal shells presented and discussed in the following section.

Starting from \eqref{eq:n_formula}, we easily find that 
\begin{equation}\label{eq:nabla_n_formula}
\nabla\vctr{n}=\frac{\nabla\vctr{m}}{\sqrt{1+\mu^2}}+
\frac{\vctr{e}_z - \mu\vctr{m}}{\sqrt{1+\mu^2}^3}\otimes\nabla\mu.
\end{equation}
Since $(\sgradn)=\nabla\n(\id-\normal\otimes\normal)$, we readily see that 
\begin{equation}\label{eq:norm_splitting}
|\sgrad\vctr{n}|^2=|\nabla \vctr{n}|^2 - |(\nabla\vctr{n})\vctr{\nu}|^2.
\end{equation}
Use of \eqref{eq:normal_representation} and \eqref{eq:norm_splitting} in lengthy, though straightforward computations finally show that
\begin{equation}\label{eq:grad_n_squared}
\begin{split}
|\sgrad\vctr{n}|^2&=
\frac{|\nabla\vctr{m}|^2}{1+\mu^2}+
\frac{|\nabla\mu|^2}{(1+\mu^2)^2}\\
&-
\frac{1}{1+|\gradh|^2}\left(
\frac{|(\nabla\vctr{m})\gradh|^2}{1+\mu^2}+
\frac{(\nabla\mu\cdot\gradh)^2}{(1+\mu^2)^2}
\right).
\end{split}
\end{equation}

The total elastic energy of a patch $\patch$ on the surface $\surface$ can now
be computed as an integral over the corresponding patch $A$ in
the $x$-$y$-plane:
\begin{equation}\label{eq:total_elatic_energy_functional}
	\mathcal{F}=
	\frac{k}{2}\int_A
	|\sgradn|^2
	\sqrt{1+|\gradh|^2}\,\dd x\dd y,
\end{equation}
where $|\sgradn|^2$ is delivered by \eqref{eq:grad_n_squared} and,
in accord with \eqref{eq:area_element}, $\sqrt{1+|\gradh|^2}$ is the Jacobian of
the transformation that lifts $A$ into $\patch$.

\section{Ellipsoidal shells}\label{sec:ellipsoids}
In this section we consider ellipsoids of revolution as a concrete example of nematic shells.
After some introductory observations, we first present equilibrium director configurations
obtained by molecular dynamics simulations performed with ellipsoids of revolution with
different aspect ratios. We then make use of the lifting method to introduce a simple model
that allows us to predict equilibrium defect locations in a closed analytic form. Using a
single fitting parameter, we find that our model is in good quantitative
agreement with the outcomes of the molecular dynamics simulations.

We assume that the surface free energy density is given in the one-constant
approximiation \eqref{eq:energy_density_one_constant}. In this case, if possible
elastic distortions are neglected, the director would prefer to align along
the principal direction of curvature that has the smallest square curvature. As a further
illustration of the lifting method, we give in Appendix \ref{app:preferred} a simple
derivation of this fact.

To find the preferred orientation on an ellipsoid of revolution, we need to
examine its principal curvatures, given in \eqref{eq:ellipsoid_curvatures}. Clearly,
both $\kappa_1$ and $\kappa_2$ are positive, so it is sufficient to look at their ratio
\begin{equation}\label{eq:kappaRatio}
\frac{\kappa_2}{\kappa_1}=1+(\eta^2-1)\frac{\rho^2}{b^2}.
\end{equation}
Here $0\leqq\rho^2/b^2\leqq 1$, and $\eta=a/b$ is the ellipsoid's aspect ratio.
On a sphere, $\eta=1$ and $\kappa_1=\kappa_2$, so there is no preferred orientation.
Furthermore,
\begin{align*}
\kappa_1&>\kappa_2\quad\text{if}\quad \eta < 1 \quad\text{(oblate)},\\
\kappa_1&<\kappa_2\quad\text{if}\quad \eta > 1 \quad\text{(prolate)}.
\end{align*}
Thus the preferred director orientation on oblate ellipsoids is along
$\vctr{e}_2$ in \eqref{eq:principalDirections}, that is along 
a parallel. The preferred director orientation on prolate ellipsoids is along
$\vctr{e}_1$ in \eqref{eq:principalDirections}, that is along 
a meridian. Equation \eqref{eq:ellipsoid_curvatures} also shows that on oblate ellipsoids
the largest curvatures are found at the equator, and the smallest curvatures are found at the poles.
The situation is reversed on prolate ellipsoids.

We thus see that the preferred orientation of the director on an ellipsoid of revolution 
is determined only by the ellipsoid's aspect ratio, independent of the position on the ellipsoid.
However, we can expect actual equilibrium director fields to follow this preference only
partly: both a director field aligned everywhere along meridians and one
aligned everywhere along parallels would feature point defects of strength one at
the poles.

\subsection{Molecular Dynamics Simulations}

We performed on ellipsoids of revolution molecular dynamics simulations similar to those
performed on spheres and reported in \cite{mirantsev:geodesic}.
The nematic shell is a thin layer of liquid crystal molecules free to glide and rotate
between two solid layers consisting of fixed molecules, which provide an effective degenerate
planar anchoring to the liquid crystal molecules, as described below. 

The interaction potential between two molecules with orientations $\vctr{\ell}_1$ and $\vctr{\ell}_2$
and with a distance $r_{12}$ between their centers of mass is \cite{luckhurst:computer_IV}
\begin{equation}\label{eq:potential}
V= V_\mathrm{iso}(r_{12})+V_\mathrm{aniso}(r_{12},\bm{\ell}_1\cdot\bm{\ell}_2),
\end{equation}
where
\begin{equation*}
V_\mathrm{iso}(r_{12})=4\varepsilon_\mathrm{iso}\left[\left(\frac{\sigma}{r_{12}}\right)^{12}- \left(\frac{\sigma}{r_{12}}\right)^{6}\right],
\end{equation*}
\begin{equation*}\label{eq:potential_split_aniso}
V_\mathrm{aniso}(r_{12},\bm{\ell}_1\cdot\bm{\ell}_2) =-\varepsilon_\mathrm{aniso}\left[\frac32(\bm{\ell}_1\cdot\bm{\ell}_2)^2-\frac12\right] \left(\frac{\sigma}{r_{12}}\right)^{6}.
\end{equation*}
Here $\sigma$ is the characteristic range of the interaction and $\varepsilon_\mathrm{iso}$
and $\varepsilon_\mathrm{aniso}$ are the isotropic and anisotropic interaction strengths.
For $\varepsilon_\mathrm{aniso}>0$, the potential encourages the molecules 
to align parallel to one another, whereas for $\varepsilon_\mathrm{aniso}<0$ the molecules
are encouraged to align at right angles to one another.
We used $\varepsilon_\mathrm{aniso}=\varepsilon_\mathrm{iso}>0$ for the interactions between
liquid crystal molecules and $\varepsilon_\mathrm{aniso}=-20\varepsilon_\mathrm{iso}<0$
(with one and the same value of $\varepsilon_\mathrm{iso}$) for the interactions
between fixed and mobile molecules.

The centres of mass of the molecules in the solid layers were frozen in random positions with
their orientations aligned along the layer normal. The liquid crystal molecules in the nematic
shell therefore prefer to orient parallel to the local tangent plane.
The system's reduced temperature was kept constant at
$T^{*}= k_{B}T/\varepsilon_\mathrm{iso} =0.9 $, where $T$ is the absolute temperature
and $k_{B}$ is the Boltzmann constant. The value prescribed for $T^{*}$ is well below
the bulk nematic-to-isotropic transition temperature, $T^{*}_{NI} = 1.05$, obtained
for a similar model system \cite{pereira:molecular}.

Simulations were started from random distributions of molecules' centers of
mass and orientations. All simulations were run for a number of time steps necessary to
reach an equilibrium state of the system.
At each time step, the equations of motion of classical particle dynamics were
solved numerically,
and the temperature of the system
was kept constant by appropriately rescaling both translational and rotational
velocities of the particles.

\begin{figure}[h]
 \begin{center}
  \parbox[c]{4.3cm}{\includegraphics[width=4.25cm]{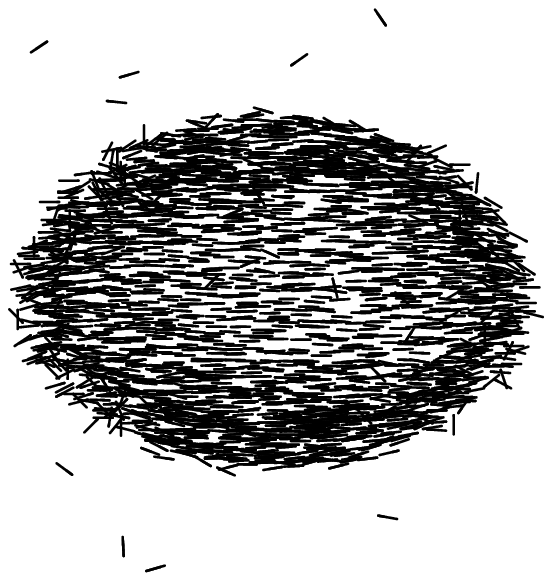}} \hfill
  \parbox[c]{3.85cm}{\includegraphics[width=3.8cm,clip]{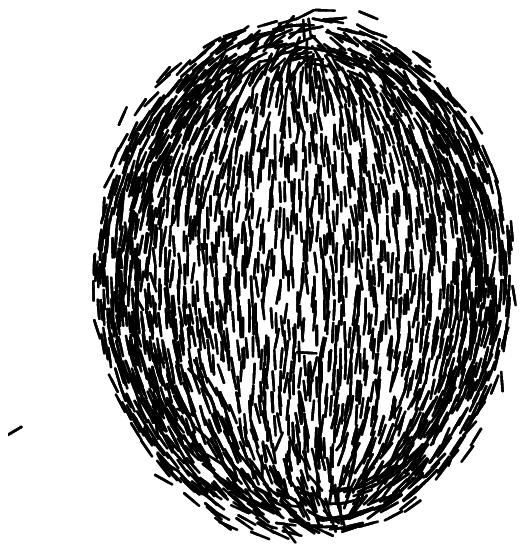}}   
 \end{center}\caption{\label{fig:sideView}Side view of ellipsoids of revolution. 
 Only molecules in the half-space facing the observer are shown. Left: $\eta=3/4$, Right: $\eta=4/3$.}
\end{figure}

\begin{figure}[h]
 \begin{center}
  \parbox[c]{4.3cm}{\includegraphics[width=4.25cm]{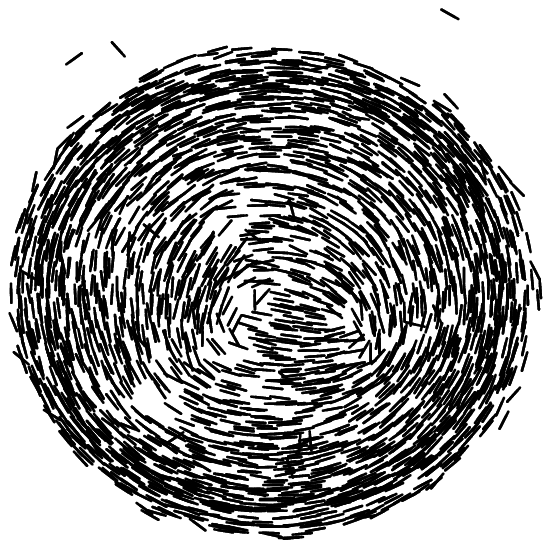}} \hfill 
  \parbox[c]{3.85cm}{\includegraphics[width=3.8cm]{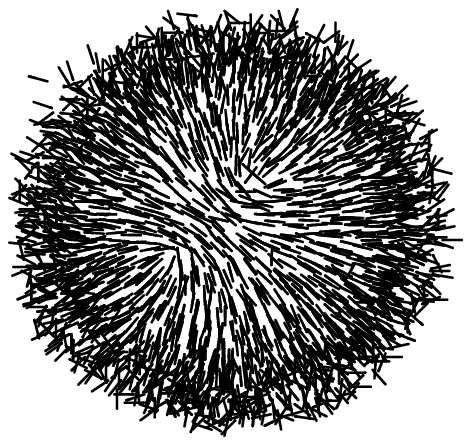}}
  \end{center}\caption{\label{fig:topView}Top view of ellipsoids of revolution. 
 Only molecules in the half-space facing the observer are shown. Left: $\eta=3/4$, Right: $\eta=4/3$.}
\end{figure}

We show in Figure \ref{fig:sideView} typical equilibrium configurations. We found, as expected, that
on oblate ellipsoids molecules predominantly align along parallels, and that on prolate ellipsoids
molecules predominantly align along meridians. However, if the same configurations are
viewed from one of the poles, Figure \ref{fig:topView}, two half-integer defects become visible.

To obtain from the molecular distribution a desciption of the local orientational order,
we introduced on the ellipsoid polar coordinates $(\phi,\Theta)$ with $\phi$ the longitude
and $\Theta$ the colatitude. At any given point $(\phi_0,\Theta_0)$ with surface normal
$\bm{\nu}_0$, we computed averages
$\langle\cdots\rangle_\mathscr{C}$ over a probing cap $\mathscr{C}$
with prescribed aperture, see Appendix \ref{app:sampling} for details.
We first computed the average second-rank tensor
\begin{equation}\label{eq:Q_k}
\mathbf{Q}=\left\langle\bm{\ell}\otimes\bm{\ell}-\frac12\mathbf{P}(\bm{\nu})\right\rangle_\mathscr{C},
\end{equation}
where $\mathbf{P}(\bm{\nu})=\mathbf{I}-\bm{\nu}\otimes\bm{\nu}$ is the projector onto
the local tangent plane.
The largest eigenvalue $\lambda$ of $\mathbf{Q}$  is the local scalar order parameter
(ranging in $[0,\frac12]$), and the corresponding normalised eigenvector of $\mathbf{Q}$
is the local director $\vctr{n}$. It can be written as 
$\vctr{n}=n_\vartheta\vctr{e}_\vartheta+n_\phi\vctr{e}_\phi+n_{\nu}\vctr{\nu}$
where $\vctr{e}_\vartheta$ is along the local meridian and $\vctr{e}_\phi$ is along the local parallel,
see \eqref{eq:e_theta_ellipsoid} and \eqref{eq:e_phi}. 

\begin{figure}[h]
 \begin{center}
  \includegraphics[width=8cm]{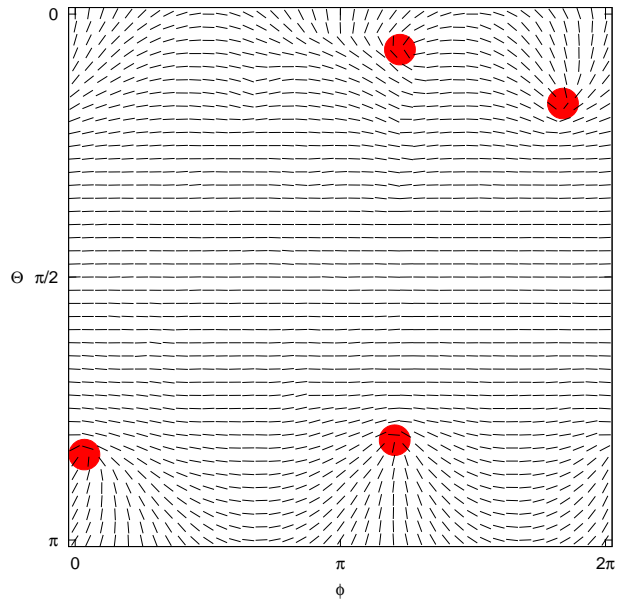}
  \end{center}\caption{\label{fig:map} Equidistant cylindrical map projection onto the $(\phi,\Theta)$-plane of
  the ellipsoid of revolution with $\eta=3/4$. The approximate defect positions are marked with discs.}
\end{figure}

For the purpose of estimating the defect distances from the poles we used
a cylindrical map projection with equidistant latitudes (and meridians) to map
the surface of an ellipsoid onto a square.\footnote{According to \cite[p.~6]{snyder:flattening}, Ptolomy credited Marinus of Tyre with the invention of this projection about 100 A.D.} While this map is neither conformal
nor area preserving, it has the obvious advantage that the defects' latitudes can
be determined by simply measuring their distances from the poles. As an example,
we reexamine the ellipsoid of revolution with
$\eta=3/4$ shown on the left in Figures \ref{fig:sideView} and \ref{fig:topView}.
We depict in Figure \ref{fig:map} in this map the projection 
$n_\vartheta\vctr{e}_\vartheta+n_\phi\vctr{e}_\phi$ of the director field onto
the local tangent plane at $(\phi,\Theta)$, see \eqref{eq:Theta_ellipsoid} for the
relationship between $\Theta$ and $\vartheta$.
Four $+1/2$ defects, two on each hemiellipsoid, are marked by circles.
Their distances from the respective closest pole where measured and the average value
was used to produce the data points used in Figure \ref{fig:comparison} below.

\subsection{Lifted Model Director Field}
It was shown in \cite{mirantsev:geodesic} how the continuum limit of the interaction potential
$V$ in \eqref{eq:potential} can be obtained by computing the average interaction energy over a geodesic circle on
the surface. One finds that
\begin{equation}\label{eq:W}
W_\mathrm{e}=\frac{K}{2}|\nabla_\mathrm{s}{\vctr{\ell}}|^2,
\end{equation}
where $K$ is a constant that depends on the surface number density and the 
radius of the geodesic circle. Our molecular dynamics simulations should therefore
correspond to a continuum model with an elastic energy in the one-constant approximation
\eqref{eq:energy_density_one_constant}.

We consider an ellipsoid of revolution with semiaxes $a$ and $b$, placed such that
its symmetry axis coincides with the $z$-axis and that its equator
lies in the $x$-$y$-plane, forming there a circle of radius $b$.
Because of the symmetry of the problem, it is sufficient to 
regard the director field as being fixed on the equator and look at just the upper
half of the ellipsoid. To represent the director field
on the hemiellipsoid, we use a single lifting map with height function $h$
given by~\eqref{eq:ellipoid_h}. However, to nondimensionalise the problem, we express
all lengths as multiples of $b$. The dimensionless height function is then given by
\begin{equation}\label{eq:dimensionlessHeight}
h(p)=\eta\sqrt{1-p^2}\, ,
\end{equation}
where $\eta=a/b$ is, as before, the ellipsoid's aspect ratio and $p=\rho/b$ measures
in the $x$-$y$-plane the distance from the origin.
The projection of the ellipsoid onto the $x$-$y$-plane is then a disc with radius $1$.
We assume that the elastic energy is given by~\eqref{eq:energy_density_one_constant}
with the norm squared of the surface gradient of the director expressed in the form
\eqref{eq:grad_n_squared}.
Our task is then to find a director field $\vctr{m}$ in the $x$-$y$-plane
that minimises the elastic energy \eqref{eq:total_elatic_energy_functional}, where
the domain of integration $A$ is the disc with radius $1$.

In principle, such a minimisation could be done numerically, but we choose here
a different approach, inspired by the director fields obtained from the molecular dynamics
simulations. As noted in Sec.~\ref{sec:lifting_tensor}, if the
director field $\vctr{n}$ on the surface is known, the correspoding field $\vctr{m}$
on the $x$-$y$-plane can be obtained as the normalised projection \eqref{eq:m_from_n_by_projection}.
This projection, albeit without the normalisation, is precisely what is
depicted in Figure \ref{fig:topView}. What we see there is the competition between
the director field near the equator, either along parallels or meridians, and
a director field near the poles with a constant projection. In between those
two fields lies a transition region with the two defects.

\begin{figure}[h]
\begin{center}
\includegraphics[width=4.1cm]{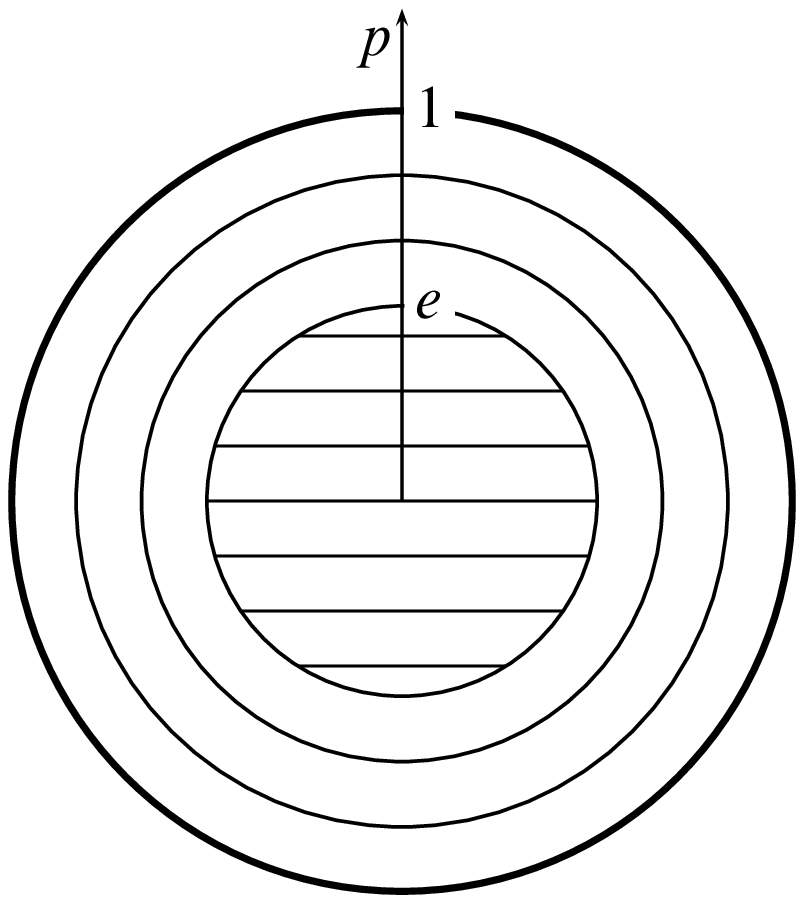}\hfill
\includegraphics[width=4.1cm]{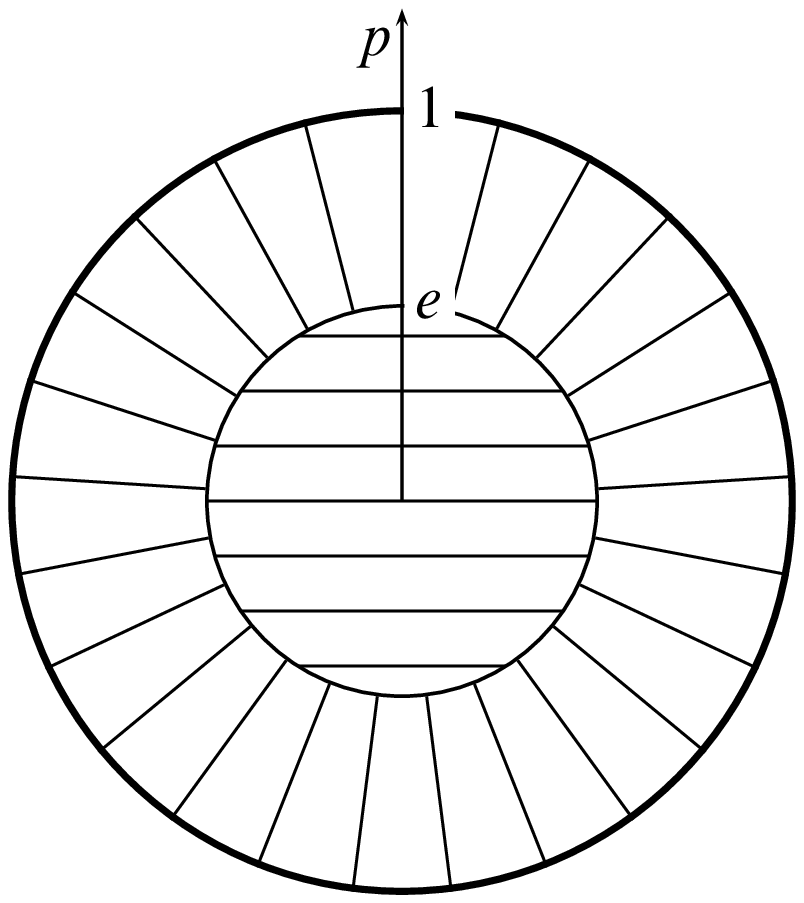}
\caption{\label{fig:patchwork} Patchwork model: 
at a distance $e$ from the origin,
a constant field borders on a circular or a radial field
that extends up to the equator at $p=1$.
Left: oblate ellipsoid, Right: prolate ellipsoid.}
\end{center}
\end{figure}

We construct a model director field in the $x$-$y$-plane as depicted
in Figure \ref{fig:patchwork}. We assume that the projections of both defects
lie at a distance $e$ from the origin, and that this is where the two competing
director fields meet. To be precise, we use
\begin{equation}
\vctr{m}=
\begin{cases}
\vctr{e}_x& 0\leqq p < e,\\
\cos\alpha\vctr{e}_\rho + \sin\alpha\vctr{e}_\phi & e < p \leqq 1,
\end{cases}
\end{equation}
with $\alpha=0$ for prolate ellipsoids and 
$\alpha=\pi/2$ for oblate ellipsoids. In a more realistic model,
two defects would be present in any such configuration, but
because they would contribute roughly the same amount to the
total energy, we simply ignore them. Across the transition line at $e$,
the director needs to perform a rotation of between $0$ and $\pi/2$.
We assume that the energy connected with this transition is proportional
to the dimensionless length of the transition line, which in turn
is proportional to $e$.

The total energy of our patchwork model thus takes the form
\begin{align}
\mathcal{F}&=
\mathcal{F}_\mathrm{po}+
\mathcal{F}_\mathrm{tr} +
\mathcal{F}_\mathrm{eq},
\end{align}
where the energy $\mathcal{F}_\mathrm{po}$ of the director field near the pole 
involves an integral in $p$ from $0$ to $e$, 
the transition energy is
\begin{equation}
\mathcal{F}_\mathrm{tr}=\lambda e 
\end{equation}
with a constant $\lambda$,
and the energy $\mathcal{F}_\mathrm{eq}$ of the director field near the equator
involves an integral in $p$ from $e$ to $1$.

There are three parameters in our model: the ellipsoid's aspect ratio $\eta$, the distance $e$ of
the projections of the defects from the origin, and the transition line energy parameter $\lambda$.
Our strategy is to find for a constant value of $\lambda$ the defect distance $e$ as a function
of $\eta$ by minimising the energy with respect to $e$, that is we solve
\begin{equation}\label{eq:equibE}
0=\frac{d\mathcal{F}_\mathrm{po}}{d e}+\lambda +\frac{d\mathcal{F}_\mathrm{eq}}{d e}
\end{equation}
for $e$.
Finally, we adjust $\lambda$ so as to best
fit the data collected from the molecular dynamics simulations.

With the dimensionless height function given by \eqref{eq:dimensionlessHeight} we
have
\begin{equation}\label{eq:gradh}
\nabla h = h'(p)\vctr{e}_\rho = \frac{-\eta p}{\sqrt{1-p^2}}\vctr{e}_\rho 
\end{equation}
and so the Jacobian of the lifting transformation is
\begin{equation}
J_L=\sqrt{1+|\nabla h|^2}=\sqrt{\frac{1+p^2(\eta^2-1)}{1-p^2}}.
\end{equation}

\paragraph{Director Near the Pole}
We use a constant field in the $x$-$y$-plane,
\begin{equation}\label{eq:mConst}
\vctr{m}=\vctr{e}_x,
\quad\text{and so}\quad
\nabla\vctr{m}=\tensor{0}. 
\end{equation}
With \eqref{eq:gradh} and $\vctr{e}_\rho=\cos\phi\vctr{e}_x+\sin\phi\vctr{e}_y$ we find
\begin{equation}\label{eq:muConst}
\mu=\vctr{m}\cdot\nabla h =\frac{-\eta p\cos\phi}{\sqrt{1-p^2}},
\end{equation}
whence
\begin{equation}\label{eq:nablaMuConst}
\nabla\mu=\frac{-\eta\cos\phi}{\sqrt{1-p^2}^3}\vctr{e}_\rho+\frac{\eta\sin\phi}{\sqrt{1-p^2}}\vctr{e}_\phi.
\end{equation}
Using \eqref{eq:mConst}, \eqref{eq:muConst}, and \eqref{eq:nablaMuConst} together with
\eqref{eq:gradh} in \eqref{eq:grad_n_squared}, we find
\begin{equation}
|\sgradn|^2=\frac{\eta^2\left\{ \cos^2\phi + (1-p^2) [1+p^2(\eta^2-1)] \sin^2\phi \right\}}{[1+p^2(\eta^2-1)] [1+p^2(\eta^2\cos^2\phi-1)]^2} .
\end{equation}
The energy $\mathcal{F}_\mathrm{po}$ between the pole and the parallel at $e$ is then
\begin{align}
\mathcal{F}_\mathrm{po}&=\frac{k}{2}\int_0^{2\pi}\int_0^e
|\sgradn|^2
J_L
p \dd p\dd\phi\\
&= \label{eq:Fpo}
\frac{1}{2}k\pi \eta^2 \int_0^e
\frac{1+[1+p^2(\eta^2-1)]^2}{(1-p^2)[1+p^2(\eta^2-1)]^2} 
p\dd p,
\end{align}
where the explicit form \eqref{eq:Fpo} is obtained by carrying out the $\phi$-integration.
The first fundamental theorem of calculus then yields
\begin{equation}\label{eq:dFpoDe}
\frac{d\mathcal{F}_\mathrm{po}}{d e}=
\frac{1}{2}k\pi \eta^2 e
\frac{1+[1+e^2(\eta^2-1)]^2}{(1-e^2)[1+e^2(\eta^2-1)]^2} .
\end{equation}

\paragraph{Director Near the Equator}
We use a field in the $x$-$y$-plane of the form
\begin{equation}\label{eq:mLox}
\vctr{m}=\cos\alpha\,\vctr{e}_\rho + \sin\alpha\,\vctr{e}_\phi,
\end{equation}
where $\alpha$ is a fixed angle. Upon lifting this field onto the
ellipsoid, we obtain
\begin{itemize}
 \item for $\alpha=0$ a director
field of {meridians}, lines of constant longitude;
 \item for $\alpha=\pi/2$ a field of {parallels},
lines of constant latitude;
 \item in general a director field whose integral lines are
{loxodromes}, lines that intersect meridians at the constant angle $\alpha$.
\end{itemize}

We have
\begin{equation}\label{eq:gradmLox}
\nabla\vctr{m}=\frac{1}{p}\left(
\cos\alpha\,\vctr{e}_\phi\otimes\vctr{e}_\phi
-
\sin\alpha\,\vctr{e}_\rho\otimes\vctr{e}_\phi
\right), 
\end{equation}
\begin{equation}\label{eq:Lox}
\mu=\vctr{m}\cdot\nabla h =\frac{-\eta p\cos\alpha}{\sqrt{1-p^2}},
\end{equation}
whence
\begin{equation}\label{eq:nablaMuLox}
\nabla\mu=\frac{-\eta\cos\alpha}{\sqrt{1-p^2}^3}\,\vctr{e}_\rho.
\end{equation}

For all values of $\alpha$ the resulting free energy density
is independent of $\phi$ so that the corresponding integration
simply yields a factor of $2\pi$:
\begin{align}\label{eq:Feq}
\mathcal{F}_\mathrm{eq}&=\frac{k}{2}\int_0^{2\pi}\int_e^1
|\sgradn|^2
J_L
p\dd p\dd\phi\\
&=
k\pi \int_e^1
|\sgradn|^2
J_L
p\dd p\label{eq:Feq2}
\end{align}

For prolate ellipsoids our patchwork model requires $\alpha=0$, which leads
to
\begin{equation}
|\sgradn|^2=\frac{p^2\eta^2+(1-p^2)[1+p^2(\eta^2-1)]^2}{p^2[1+p^2(\eta^2-1)]^3}.
\end{equation}
Using this in \eqref{eq:Feq2} and differentiating with respect to $e$ we obtain
\begin{equation}\label{eq:dFpoPDe}
\frac{d\mathcal{F}_\mathrm{eq}^\mathrm{p}}{d e}=
-k\pi\frac{e^2\eta^2+(1-e^2)[1+e^2(\eta^2-1)]^2}{e\sqrt{(1-e^2)[1+e^2(\eta^2-1)]^5}}.
\end{equation}

For oblate ellipsoids our patchwork model requires $\alpha=\pi/2$, which leads
to
\begin{equation}
|\sgradn|^2=\frac{1}{p^2},
\end{equation}
and using this in \eqref{eq:Feq2} gives
\begin{equation}\label{eq:dFpoODe}
\frac{d\mathcal{F}_\mathrm{eq}^\mathrm{o}}{d e}=
-k\pi\frac{\sqrt{1+e^2(\eta^2-1)}}{e\sqrt{1-e^2}}.
\end{equation}

\subsection{Comparison}
\begin{figure}[h]
 \begin{center}
  \includegraphics[width=8cm]{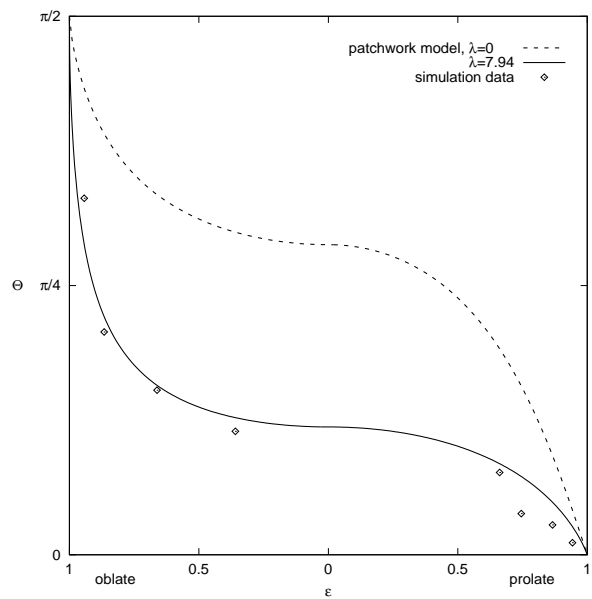}
  \end{center}\caption{\label{fig:comparison} Comparison of simulation data and analytical model.
  The dashed line corresponds to zero transition energy, the solid line was obtained
  by a least-square fit of the phenomenological transition line energy constant $\lambda$.}
\end{figure}

We used in \eqref{eq:equibE} the expression \eqref{eq:dFpoDe} together with
\eqref{eq:dFpoPDe} for $\eta>1$ and 
\eqref{eq:dFpoODe} for $\eta<1$. They result eventually in a polynomial
equation for $e$, which was solved for fixed $\lambda$ numerically
with $e=1/2$ as starting value for $200$ values of $\eta$. The outcome
is shown in Figure \ref{fig:comparison}. To obtain a finite range on
the abscissa, we used in the figure instead of the aspect ratio $\eta$
the excentricity $\epsilon$, given by
\begin{equation}
\epsilon=
\begin{cases}
\sqrt{1-\eta^2}& \eta\leqq 1,\\[2pt]
\sqrt{1-\eta^{-2}} &\eta >1.
\end{cases}
\end{equation}
The ordinate shows the polar angle $\Theta$ of the defect position, given by
\begin{equation}
\Theta=\arctan \frac{e}{h(e)}=\arctan\frac{e}{\eta\sqrt{1-e^2}}.
\end{equation}

Even when the transition between the two model director fields around the pole and equator is
completely ignored, $\lambda=0$, our patchwork model captures in a qualitatively correct way the effect of
the ellipsoids' shape on the defect postion: the more oblate an ellipsoid, the closer the
defects are to the equator, and the more prolate an ellipsoid, the closer the defects are to the poles.

The transition line energy basically penalises closeness of defects to the equator, and its
net effect in the model is to push the transition line towards the pole. With the value
$\lambda=7.94$, obtained by a least-square fit, our model shows good quantitative agreement with
the molecular dynamics simulation data.

\section{Conclusions}\label{sec:conclusions}
The main objective of this paper is to propose a systematic method to represent order and its distortions on curved material surfaces by reading them off from a flat, reference surface. Clearly, variational problems staged on generally curved surfaces, although graphs of an appropriate height function, remain difficult to solve, but incorporating the geometric details into the functional form of the energy may be computationally advantageous, as shown in the applications to nematic shells presented in Secs.~\ref{sec:shells} and \ref{sec:ellipsoids}.

Our method is sufficiently general to allow for a surface differential calculus somewhat more agile than the traditional approach based on an atlas of local coordinate maps. The main mathematical tool employed here is the lifting tensor $\Lift$, which converts a planar director field $\m$ into a surface tangential director field $\n$. Although, in principle, the curved surface $\surface$ treated by our method may well be flexible, the director field $\m$ lifted into the actual order descriptor $\n$ is just a formal artifice to represent $\n$, precisely as is the flat projection $S$ of $\surface$. In our approach, whereas $\n$ is the lifted image of $\m$, the latter is not generally \emph{imprinted} in the flat surface $S$, precisely as $\surface$ is not generally the material image of $S$ under deformation.

When the actual deformation of $S$ into $\surface$, here replaced by the height-function parameterization,
is an important ingredient of the theory, as is the case for glassy and elastomeric
nematics~\cite{mostajeran:curvature,mostajeran:encoding}, our lifting tensor $\Lift$ fails to capture the entire
richness in mechanical behaviours exhibited by these systems. In particular, external stimuli brought about by changes
in either temperature or illumination prescribe the principal stretches of an initially flat nematic glassy sheet
along the imprinted nematic director $\m$ and the direction orthogonal to that. Describing the deformation
undergone by a flexible nematic sheet under the kinematic constraints imposed by the external stimuli and
physical anchoring is a challenge that requires extending the notion of lifting tensor introduced in this paper,
so as to keep track of how material body points are carried with their order parameters from $S$ over to $\surface$.
Such an extension, which is currently underway, features an in-plane gliding component of the deformation that supplements the elevation described by the height function. We trust that a new method could be available in the future to describe both the distortion of imprinted order tensors and the deformation of their material substrates.


\appendix
\section{Perferred Orientation}\label{app:preferred}
We want to find the preferred orientation of the director
in the one-constant approximation \eqref{eq:energy_density_one_constant}
on a surface at a point where the principal curvatures
are $\kappa_1$ and $\kappa_2$. We choose coordinates such
that the point is the origin, the tangent plane at the point is the $x$-$y$-plane,
and the principal directions of curvature are $\vctr{e}_x$ and $\vctr{e}_y$.
The curvature tensor $\tensor{H}$ is thus
\begin{equation}
\tensor{H}=\kappa_1\vctr{e}_x\otimes\vctr{e}_x +\kappa_2\vctr{e}_y\otimes\vctr{e}_y.
\end{equation}

The height of the surface over the $x$-$y$-plane at a point
with position vector $\vctr{r}=x\vctr{e}_x+y\vctr{e}_y$ is then given by Taylor's theorem as
\begin{align}
h(\vctr{r})&= h(\vctr{0}) + \nabla h(\vctr{0})\cdot\vctr{r} +
\frac{1}{2} \vctr{r}\cdot[\nabla^2 h(\vctr{0})]\vctr{r}+o(|\vctr{r}|^2)\\
&=\frac{1}{2}\vctr{r}\cdot\tensor{H}\vctr{r}+o(|\vctr{r}|^2)
\end{align}
because, with our choice of coordinates, $h(\vctr{0})=0$, $\nabla h(\vctr{0})=\vctr{0}$,
and the Hessian $\nabla^2 h(\vctr{0})$ is equal to the curvature tensor
$\tensor{H}$, see, for example, \cite[p.137]{spivak:diffferential3} or \cite[\S3.3]{carmo:differential}.
It follows that
\begin{align}\label{eq:hofr}
\nabla h=\tensor{H}\vctr{r}+o(|\vctr{r}|)=x\kappa_1\vctr{e}_x+y\kappa_2\vctr{e}_y+o(|\vctr{r}|).
\end{align}

We now consider a constant director field in the $x$-$y$-plane,
\begin{equation}\label{eq:constM}
\vctr{m}=\cos\alpha\, \vctr{e}_x + \sin\alpha\,\vctr{e}_y,
\end{equation}
and we want to determine the angle $\alpha$ for which the free energy density
at the origin is minimal. We have $\nabla\vctr{m}=\tensor{0}$ throughout,
and at the orgin $\nabla h=\vctr{0}$. Equation \eqref{eq:grad_n_squared} at the origin 
therefore simplifies to
\begin{equation}
|\sgrad\vctr{n}|^2=\frac{|\nabla\mu|^2}{(1+\mu^2)^2}=|\nabla\mu|^2\qquad\text{with }\mu=\vctr{m}\cdot\nabla h.
\end{equation}
With \eqref{eq:hofr} and \eqref{eq:constM}, we find
$\mu=x\kappa_1\cos\alpha+y\kappa_2\sin\alpha+o(|\vctr{r}|)$ and so  
$\nabla\mu=\kappa_1\cos\alpha\,\vctr{e}_x+\kappa_2\sin\alpha\,\vctr{e}_y+o(1)$.
Thus at the origin we have
\begin{equation}
|\sgrad\vctr{n}|^2=\kappa_1^2\cos^2\alpha+\kappa_2^2\sin^2\alpha.
\end{equation}

The free energy density at the origin is hence proportional to a function $f$ of the
director angle $\alpha$ given by
\begin{equation}
f(\alpha)=\kappa_1^2\cos^2\alpha+\kappa_2^2\sin^2\alpha ,
\end{equation}
and so
\begin{equation}
f'(\alpha)=(\kappa_2^2-\kappa_1^2)\sin 2\alpha.
\end{equation}
The minimum free energy density is obtained for
\begin{align}
\alpha=0,\ \vctr{m}=\vctr{e}_x\quad\text{if}\quad \kappa_1^2<\kappa_2^2, \\
\alpha=\frac{\pi}{2},\ \vctr{m}=\vctr{e}_y\quad\text{if}\quad \kappa_2^2<\kappa_1^2.
\end{align}
The director prefers to align along the direction of smallest square curvature.

\section{Sampling Over Axisymmetric Surfaces}\label{app:sampling}
An axisymmetric surface $\surface$ can also be represented by two scalar functions, $\rho(\vt)$ and $z(\vt)$,
which parameterize the planar curve whose revolution (about $\ez$) generates $\surface$.
Relative to a Cartesian frame $(\ex,\ey,\ez)$ with origin in $o$, a point $p$ in $\surface$ is identified by the vector
\begin{align}\label{eq:radial}
\radial(\vt,\vp)&=p(\vt,\vp)-o\nonumber \\
&=\rho(\vt)(\cos\vp\,\ex+\sin\vp\,\ey)+z(\vt)\ez,
\end{align}
where $\vt\in[0,\pi]$ and $\vp\in[0,2\pi]$. Conventionally, we call North and South poles the points at $\vt=0$ and $\vt=\pi$, respectively. In general, the angle $\vt$ differs from the \emph{polar} angle $\Theta$ relative to the axis $\ez$, which is given by
\begin{equation}\label{eq:Theta_angle}
\Theta=\arctan\left(\frac{\rho(\vt)}{z(\vt)}\right).
\end{equation}

The radial unit vector in the $x$-$y$-plane  is denoted by
\begin{equation}\label{eq:e_rho}
\er:=\cos\vp\,\ex+\sin\vp\,\ey,
\end{equation}
while the azimuthal unit vector orthogonal to $\er$ in the $x$-$y$-plane is delivered by
\begin{equation}\label{eq:e_phi}
\ep:=-\sin\vp\,\ex+\cos\vp\,\ey.
\end{equation}
At a point $p(\vt,\vp)$ on $\surface$, the unit tangent vector $\et$ to the local meridian oriented along the direction of increasing $\vt$ is given by
\begin{equation}\label{eq:e_theta}
\et=\frac{1}{\sqrt{\rho'^2+z'^2}}\left[\rho'(\cos\vp\,\ex+\sin\vp\,\ey)+z'\ez\right],
\end{equation}
where a prime $'$ denotes differentiation with respect to $\vt$. The unit outer normal $\normal:=\et\times\ep$ is accordingly given by
\begin{equation}\label{eq:normal}
\normal=\frac{1}{\sqrt{\rho'^2+z'^2}}\left[-z'(\cos\vp\,\ex+\sin\vp\,\ey)+\rho'\ez\right].
\end{equation}

A \emph{crust} of thickness $d$ above the surface $\surface$ is bounded by the surface $\surface_d$ represented by
\begin{equation}\label{eq:radial_d}
\radial_d(\vt,\vp):=\radial+d\normal,
\end{equation}
where $\radial$ is as in \eqref{eq:radial} and $\normal$ as in \eqref{eq:normal}.

The curvature tensor $\sgrad\normal$ of $\surface$ can also be described in the local frame $(\et,\ep,\normal)$ by use of the parameterization \eqref{eq:radial}; we readily obtain a formula that reminds us of \eqref{eq:principal_basis_representation},
\begin{equation}\label{eq:curvature_tensor}
\sgrad\normal=\frac{z'\rho''-\rho'z''}{(\rho'^2+z'^2)^{3/2}}\et\otimes\et -\frac{z'}{\rho\sqrt{\rho'^2+z'^2}}\ep\otimes\ep.
\end{equation}
It follows from \eqref{eq:curvature_tensor}
that the principal curvatures $\kappa_\vt$ and $\kappa_\vp$ along the principal curvature directions $\et$ and $\ep$ as
\begin{subequations}\label{eq:principal_curvatures}
	\begin{align}
	\kappa_\vt&=\frac{z'\rho''-\rho'z''}{(\rho'^2+z'^2)^{3/2}},\label{eq:sigma_theta}\\
	\kappa_\vp&=-\frac{z'}{\rho\sqrt{\rho'^2+z'^2}},\label{eq:sigma_phi}
	\end{align}
\end{subequations}
which provide expressions alternative, but equivalent to those in \eqref{eq:axis_symmetric_curvatures}, once we identify $\e_1$ with $\et$ and $\e_2$ with $\ep$, respectively.

A sampling area on $\surface$ around the point $p(\vt_0,\vp_0)$ can be identified as the collection of
all points $p(\vt,\vp)$ in one and the same connected component\footnote{Such a proviso is necessary for a non-convex surface $\surface$.} with $p(\vt_0,\vp_0)$,
such that the normal $\normal$ lies within a cone of semi-amplitude $\alpha_0$ around
the normal $\normal_0$ at $p(\vt_0,\vp_0)$. Formally, this requirement is embodied by the inequality
\begin{equation}\label{eq:sampling_area}
\frac{1}{\sqrt{\rho_0'^2+z_0'^2}}\frac{1}{\sqrt{\rho'^2+z'^2}}\left[\cos(\vp-\vp_0)z'z_0'+\rho'\rho_0'\right] \geqq\cos\alpha_0,
\end{equation}
where $\rho_0'$ and $z'_0$ are shorthands for $\rho'(\vt_0)$ and $z'(\vt_0)$, respectively.

For an \emph{ellipsoid} of revolution with semi-axes $a$ and $b$, along $\ez$ and $\er$, respectively, the functions $\rho$ and $z$ are given by
\begin{equation}\label{eq:rho_z_ellipsoid}
\rho(\vartheta)=b\sin\vt,\qquad z(\vt)=a\cos\vt.
\end{equation}
By using these functions in \eqref{eq:radial}, \eqref{eq:e_theta}, \eqref{eq:normal}, \eqref{eq:principal_curvatures}, and \eqref{eq:sampling_area}, we arrive at the following formulae:
\begin{subequations}
	\begin{equation}\label{eq:radial_ellipsoid}
	\radial(\vt,\vp)=b\sin\vt(\cos\vp\,\ex+\sin\vp\,\ey)+a\cos\vt\,\ez,
	\end{equation}
	\begin{align}\label{eq:e_theta_ellipsoid}
	\et&=\frac{1}{\sqrt{\cos^2\vt+\eta^2\sin^2\vt}}\nonumber\\
	&\times\left[\cos\vt(\cos\vp\,\ex+\sin\vp\,\ey) -\eta\sin\vt\,\ez\right],
	\end{align}
	\begin{align}\label{eq:normal_ellipsoid}
	\normal&=\frac{1}{\sqrt{\cos^2\vt+\eta^2\sin^2\vt}}\nonumber\\
	&\times\left[\eta\sin\vt(\cos\vp\,\ex+\sin\vp\,\ey) +\cos\vt\,\ez\right],
	\end{align}
	\begin{equation}\label{eq:sigma_theta_ellipsoid}
	\sigma_\vt=\frac{1}{b}\frac{\eta}{(\cos^2\vt+\eta^2\sin^2\vt)^{3/2}},
	\end{equation}
	\begin{equation}\label{eq:sigma_phi_ellipsoid}
	\sigma_\vp=\frac{1}{b}\frac{\eta}{\sqrt{\cos^2\vt+\eta^2\sin^2\vt}},
	\end{equation}
	\begin{align}\label{eq:sampling_area_ellipsoid}
	&\frac{1}{\sqrt{\cos^2\vt_0+\eta^2\sin^2\vt_0}}\frac{1}{\sqrt{\cos^2\vt+\eta^2\sin^2\vt}}\nonumber\\ &\times[\eta^2\cos(\vp-\vp_0)\sin\vt\sin\vt_0+\cos\vt\cos\vt_0 ]\nonumber\\
	&\geqq\cos\alpha_0,
	\end{align}
\end{subequations}
where $\eta:=a/b$ is the ellipsoid's aspect ratio.
It is also easily checked with the aid of \eqref{eq:Theta_angle} that for an ellipsoid the polar angle $\Theta$ is related to the angle $\vt$ through
\begin{equation}\label{eq:Theta_ellipsoid}
\Theta=\arctan\left(\frac{1}{\eta}\tan\vt\right).
\end{equation}

In the local frame $(\et,\ep,\normal)$, the molecular director $\bm{\ell}$ is represented by
\begin{equation}\label{eq:molecular_director}
\mol=\ell_\vt\et+\ell_\vp\ep+\ell_\nu\normal.
\end{equation}

To compute averages at a given point  $(\phi_0,\Theta_0)$ with surface normal
$\bm{\nu}_0$, we used the criterion \eqref{eq:sampling_area_ellipsoid} to include
all molecules found at positions where the surface normal $\bm{\nu}$ deviated by less than a
specified angle $\alpha_0$ from $\bm{\nu}_0$. Using a fixed angle for the averaging
produced poor results for ellipsoids of revolution with large excentricities, either at the
poles or at the equator. Rather than attempting to scale the angle $\alpha_0$ using the
local surface area of the ellipsoid, we used the heuristic formula
\begin{equation}\label{eq:capScaling}
\alpha_0=\frac{\alpha\eta}{\cos^2\vartheta_0+\eta^2\sin^2\vartheta_0} 
\end{equation}
with $\alpha=6^\circ$. The effect of \eqref{eq:capScaling} is to scale the
cap size by $\eta$ at the poles and by $1/\eta$ at the equator, which produces
the desired effect both for prolate and oblate ellipsoids of revolution.


\end{document}